\documentclass{article}

\usepackage{arxiv}

\usepackage[utf8]{inputenc} 
\usepackage[T1]{fontenc}    
\usepackage{hyperref}       
\usepackage{url}            
\usepackage{booktabs}       
\usepackage{amsfonts}       
\usepackage{nicefrac}       
\usepackage{microtype}      
\usepackage{lipsum}
\usepackage{graphicx}
\usepackage{upgreek}
\usepackage{stfloats}
\usepackage{amsmath}

\title{Design of a high throughput telescope based on scanning off-axis Three-Mirror Anastigmat system}

\author{
   Huiru Ji, Zhengbo Zhu, Hao Tan, Yuefan Shan, Wei Tan, Donglin Ma\thanks{Shenzhen Huazhong University of Science and Technology, Shenzhen 518057, China}   \\
  School of Optical and Electronic Information and Wuhan National Laboratory of Opto-electronics\\
  Huazhong University of Science and Technology\\
  Wuhan 430074, China \\
  \texttt{madonglin@hust.edu.cn} \\
  }

\begin{document}
\maketitle

\begin{abstract}
High throughput optical system is defined to possess the features of both large field of view (FOV) and high resolution. However, it is full of challenge to design such a telescope with the two conflicting specifications at the same time. In this paper, we propose a method to design a high throughput telescope based on the classical off-axis Three-Mirror Anastigmat (TMA) configuration by introducing a scanning mechanism. We derive the optimum initial design for the TMA system with no primary aberrations through characteristic ray tracing. During the design process, a real exit pupil is necessitated to accommodate the scanning mirror. By gradually increasing the system’s FOV during the optimization procedure, we finally obtained a high throughput telescope design with an F-number of 6, a FOV of 60$^{\circ}$×1.5$^{\circ}$, and a long focal length of 876mm. In addition, the tolerance analysis is also conducted to demonstrate the instrumentation feasibility. We believe that this kind of large rectangle FOV telescope with high resolution has broad future applications in the  optical remote sensing field.
\end{abstract}

\keywords{High Throughput Telescope\and Off-axis TMA\and Scanning Mechanism\and Large FOV\and High Resolution }

\section{Introduction}

Driven by the rapid development of remote sensing, the goal of optical systems design in space exploration, astronomical imaging, earth science, military application and many other fields, is moving towards large field of view (FOV) and high resolutions. The researchers mainly concern about the high throughput telescopes, such as Sky Mapper Wide Field Telescope \cite{rakich2006skymapper}, Panoramic Survey Telescope and Rapid Response System (Pan-STARRS) \cite{bolden2010panoramic}, VST Telescope \cite{schipani2012vst}, the visible and infrared survey telescope for astronomy (VISTA) \cite{sutherland2015visible}, the Javalambre Survey Telescope (JST) \cite{cenarro2015javalambre}, which are mainly designed for remote sensing while maintaining high image quality. However, as limited by the size of the detectors and the structure of the optical systems, the FOV of these telescopes is usually very small as it is contradictory to long-distance imaging. Taking the famous Large Synoptic Survey Telescope (LSST) as an example, this large coaxial  Three-Mirror Anastigmat (TMA) telescope is composed of an 8.4 meters primary mirror, a 3.4 meters secondary mirror and a 5 meters tertiary mirror, has a FOV of 3.5$^{\circ}$ \cite{sebag2012lsst}. Correspondingly, the large FOV systems usually cannot achieve long-distance imaging as well \cite{wu2019optical}. 

In the past few decades, a great deal of researches have been done to balance this contradiction and have made some achievements. As a promising alternative, off-axis TMA systems have attracted the researchers' attention because of the incomparable advantages of compactness, no central obstruction, high degree of freedoms and free of chromatic aberration. These features have great potential to provide good optical solutions to meet the requirements of high resolution while maintaining a wide FOV \cite{li2012optical,zhu2014design,meng2019design}. The widely applied off-axis configurations usually make the aperture stop offset \cite{bai2019aberrational} or FOV biased \cite{zhang2012design} to avoid obscuration and achieve large FOV. Consequently, the non-rotationally symmetric aberrations dominated by astigmatism and coma \cite{thompson2005description} caused by the off-axis structures are difficult to eliminate and make the manufacturing as well as the alignment a big challenge. Fortunately, with the development of aberration theory, optical system optimization mechanism and the manufacturing technology, the application of freeform surfaces has shown great potential in the off-axis optical systems \cite{fuerschbach2014assembly,yang2015direct,zhong2017initial,muslimov2017combining,gu2020optical}. This is owing to that the freeform surfaces possess the feature of non-rotational symmetry,
which can reduce the asymmetric aberration to a certain extent caused by the off-axis of the system \cite{shi2016analysis,ju2018aberration}.

\begin{figure}[ht!]
\centering
\includegraphics[width=.8\textwidth]{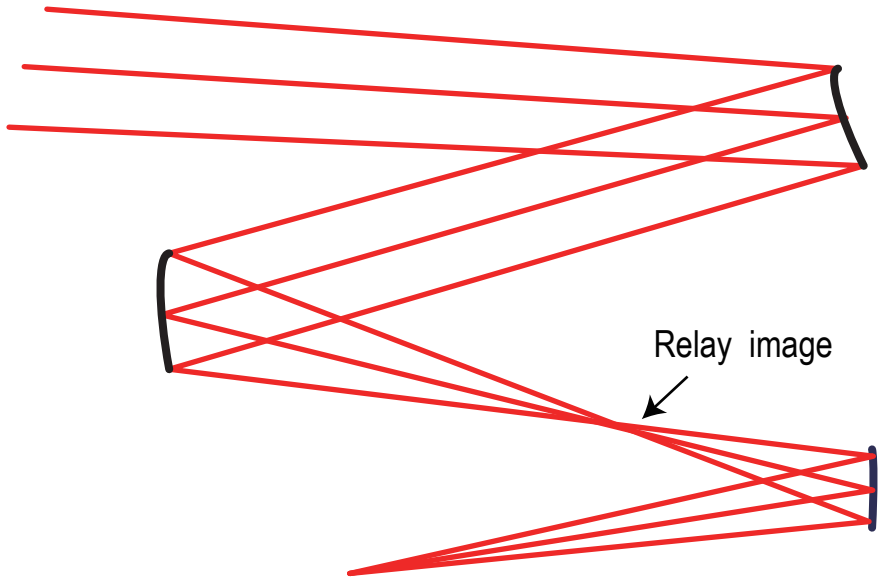}
\caption{ Schematic diagram of off-axis TMA system with a relay image.}
\label{fig:1}
\end{figure}

In this paper, we propose a design method for high throughput telescope with TMA configuration based on coaxial Sedel aberration theory. The design is aimed to achieve a FOV of 60$^{\circ}$×1.5$^{\circ}$ and make image performance close to the diffraction limit for remote sensing. For the requirement of high resolution, an off-axis TMA structure with a relay image is adopted as the optical configuration as illustrated in Fig. \ref{fig:1}. We place a scanning mirror at the real exit pupil to guide the light rays coming from the different FOVs to the overlapped position on the image plane, thus reducing the size of the image plane to fit the detector. By adjusting the rotation angle of the scanning mirror, the images of various FOVs detected will be stitched together to achieve a large full FOV. When the scanning mirror is fixed in a certain spatial state, the instantaneous FOV is small which means a high resolution. This working mechanism solves the contradiction between large FOV and high quality remote imaging simultaneously.

The whole design process begins with deriving the optimum coaxial initial structure with no primary aberrations through ray tracing of two characteristic rays, which is detailed in section \ref{sec:2}. In section \ref{sec:3}, a progressive optimization is implemented to gradually increase the optical performance. Specifically, we successively add the off-axis quantity to improve FOV and design a scanning mirror to improve the optical resolution. Then, we have analyzed the tolerance of the obtained optical system to demonstrate the instrumentation feasibility in section \ref{sec:4}. Finally, a brief summary is given in section \ref{sec:5}.

\section{Design principle and initial structure determination via third order Seidel aberration}
\label{sec:2}

\subsection{Design Specifications and Analysis}

The design purpose and specifications are expressly listed in Table \ref{tab:1}. Diffraction limited image quality means that the RMS size of the image spots should be smaller than the system’s Airy disk radius, which is expressed by
\begin{equation}
R = \frac{{1.22\lambda }}{D}f{\rm{ = }}1.22\lambda  \cdot {F \mathord{\left
/{\vphantom {F \# }} \right.
\kern-\nulldelimiterspace} \# },
\label{eq:1}
\end{equation}
where $D$ denotes the entrance pupil diameter of the system, $f$ represents the effective focal length (EFL), and $F/\#$ is the F number. For instance, if the working wavelength equals 10$\upmu$m, $R$ is calculated as 73.2$\upmu$m.
\begin{table}[htbp]
\centering
\caption{ Design Specifications}
\begin{tabular}{lll}
\hline
F number & 6 \\
FOV & 60$^{\circ}$×1.5$^{\circ}$ \\
Entrance pupil diameter & 160mm \\
Wavelength range & 8$\sim$12$\upmu$m\\
Image quality & Reach diffraction limit\\
Detector pixel size & 50$\upmu$m×50$\upmu$m\\
Detector size & 640mm×512mm\\
\hline
\end{tabular}
  \label{tab:1}
\end{table}

\begin{figure}[ht!]
\centering
\includegraphics[width=.8\textwidth]{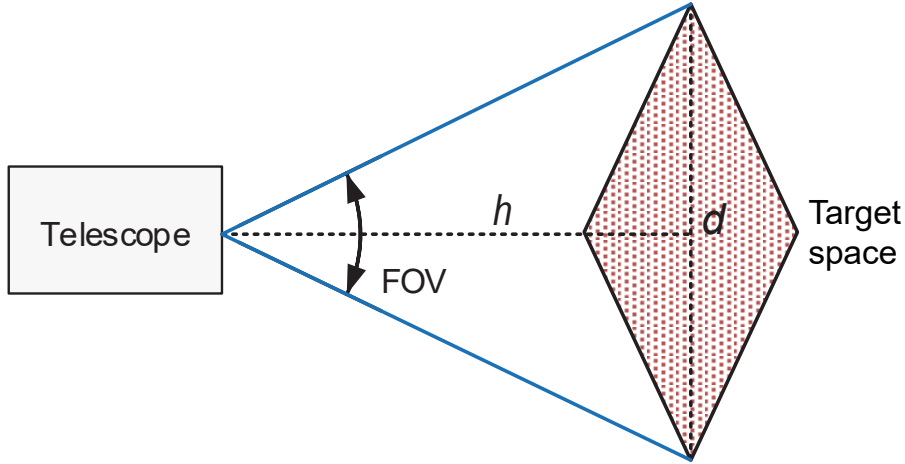}
\caption{ Geometrical relationship between FOV and detection range $d$
as well as the detection distance $h$.}
\label{fig:2}
\end{figure}

For remote optical sensing applications, a larger FOV is usually preferred, which means a wider target space, as illustrated in Fig. \ref{fig:2}. For a two-dimension case, the target dimension for a specific FOV can be expressed as:
\begin{equation}
d = 2h \cdot \tan \frac{{{\rm{FOV}}}}{2},
\label{eq:2}
\end{equation}
where $d$ represents the width of the target that can be observed by the telescope and $h$ stands for the detection distance. So, when $h$ is fixed, an optical system with a wider FOV leads to a larger $d$ and needs a higher optical resolution. Also, this will results in a larger size of the mirrors and increasing the difficulty of manufacturing and alignment. To solve this problem, some studies have proposed the use of FOV segmentation \cite{jahn2017innovative} or the push broom scanner \cite{risse2008novel} to divide the large FOV into a superposition of small sub-FOVs. As a rule of thumb, a scanning mirror is also adopted in this paper to divide the full FOV into several sub-FOVs with the usage of multi-configuration. For each sub-FOV, a high optical resolution is guaranteed.

\subsection{Design of the Initial Structure}
For the off-axis TMA systems with a relay image, there are two types of structural designs that are widely adopted according to the optical power distribution of mirrors, namely “concave-convex-concave” systems \cite{meng2016off}  and “convex-concave-concave” systems\cite{zhang2012design}.
As a rule of thumb, the latter one has a relative stronger ability to correct spherical aberration, coma, astigmatism and field curvature compared with the former one, and is chosen as the optical configuration in this paper.

\begin{figure}[ht!]
\centering
\includegraphics[width=.8\textwidth]{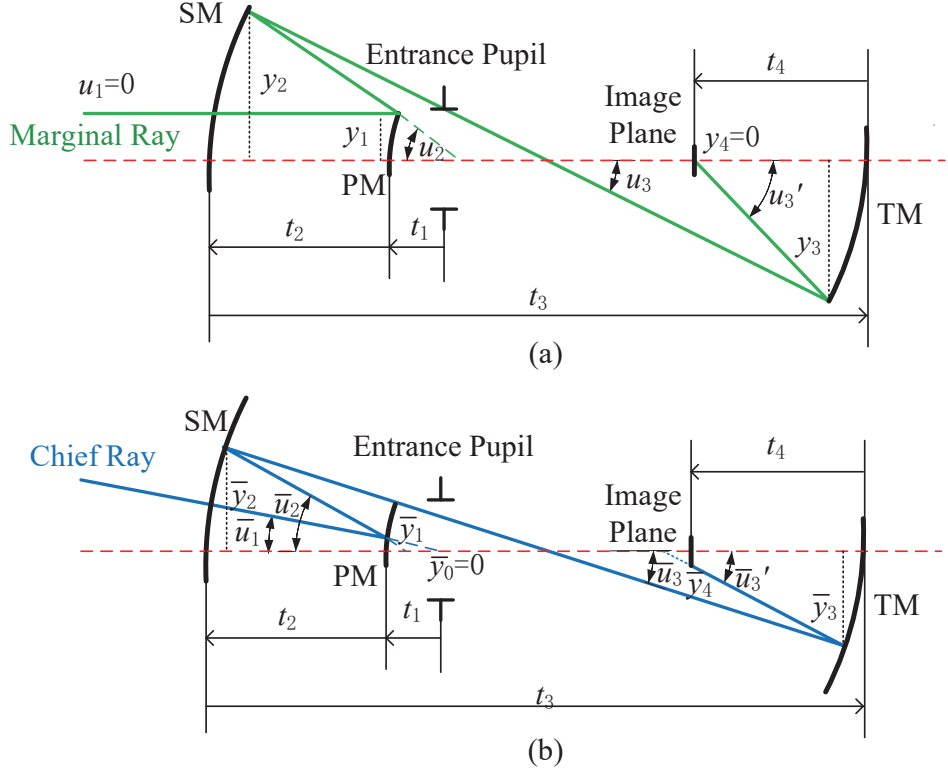}
\caption{ Ray tracing of initial coaxial TMA structure: (a) trace a marginal ray; (b) trace a chief ray.
}
\label{fig:3}
\end{figure}

We start the initial design by building a coaxial structure which includes a relay image and a real exit pupil, as shown in Fig. \ref{fig:3}. The aperture stop of system is behind the prime mirror (PM), and all mirrors are designed as spherical surfaces. $t_{1}$, $t_{2}$, $t_{3}$ and $t_{4}$ represent the distance between the entrance pupil and PM, the distance between PM and the secondary mirror (SM), the distance between SM and the tertiary mirror (TM), and the distance between TM and the image plane respectively. $r_{1}$, $r_{2}$, and $r_{3}$ denote the curvature radii of PM, SM, and TM, respectively. We set transmission media refractive index as $n_{1}=n'_{2}=n_{3}=1$, $n'_{1}=n_{2}=n'_{3}=-1$.

 For the  propagation of the marginal ray as illustrated in Fig. \ref{fig:3} (a), $u$ is the paraxial ray's incident slope  with respect to the optical axis, and $u'$ is the corresponding exit slope. Marginal rays enter PM in a parallel way with respect to the optical axis, so $u_1=0$. $y_1$, $y_2$, and $y_3$  represent the heights of the marginal ray on PM, SM, TM in turn. For PM, the optical power can be determined as:
\begin{equation}
{\phi _1}{\rm{ = }}\left( {{{n'}_1} - {n_1}} \right) \cdot {c_1} =  - \frac{2}{{{r_1}}},
\label{eq:3}
\end{equation}
where $c_1$ is the curvature of the PM. Similarly, for SM and TM, we have
\begin{equation}
\left\{ \begin{array}{l}
{\phi _2} = \frac{2}{{{r_2}}}\\
{\phi _3} =  - \frac{2}{{{r_3}}}.
\end{array} \right.
\label{eq:4}
\end{equation}

The relationship between the ray height and the ray slope on different mirrors can be determined by the paraxial ray tracing formulas, which are expressed as.
\begin{equation}
\left\{ \begin{array}{l}
{{n'}_i}{{u'}_i} = {n_i}{u_i} - {y_i}{\phi _i}\\
{y_{i + 1}} = {y_i} + {{u'}_i}{{t'}_i}\\
{{t'}_i} = {t_{i + 1}}\\
{{u'}_i} = {u_{i + 1}},
\end{array} \right.\
\label{eq:5}
\end{equation}
where $i (i = 1, 2, 3, 4)$ represents PM, SM, TM and image plane respectively. The ray tracing calculation method of the chief ray is similar to that of the marginal ray, as shown in Fig. \ref{fig:3}(b). The extension line of the incident ray passes through the center of the entrance pupil, which is recorded as the 0th surface, so $\overline{y}_0$\ = 0. The ray tracing follows:
\begin{equation}
\left\{ \begin{array}{l}
{{n'}_i}{{\bar u'}_i} = {n_i}{{\bar u}_i} - {{\bar y}_i}{\phi _i}\\
{{\bar y}_{i + 1}} = {{\bar y}_i} + {{\bar u'}_i}{{t'}_i}\\
{{t'}_i} = {t_{i + 1}}\\
{{\bar u'}_i} = {{\bar u}_{i + 1}}.
\end{array} \right.\
\label{eq:6}
\end{equation}

According to Seidel aberration theory, in a rotationally symmetric optical system, five monochromatic aberrations including spherical aberration ($S_{I}$), coma ($S_{II}$), astigmatism ($S_{III}$), Petzval curvature of field ($S_{IV}$) and distortion ($S_{V}$) can be expressed by the structural parameters of the system:
\begin{equation}
\left\{ {\begin{array}{*{20}{l}}
{{S_I} =  - \sum {{A^2} \cdot y \cdot \Delta \left( {\frac{u}{n}} \right)} }\\
{{S_{II}} =  - \sum {\bar AA \cdot y \cdot \Delta \left( {\frac{u}{n}} \right)} }\\
{{S_{III}} =  - \sum {{{\overline A }^2} \cdot y \cdot \Delta \left( {\frac{u}{n}} \right)} }\\
{S_{IV}} =  - \sum {{H^2} \cdot c \cdot \Delta \left( {\frac{1}{n}} \right)} \\
{S_V} =  - \sum {\left\{ {\frac{{{{\overline A }^3}}}{A} \cdot y \cdot \Delta \left( {\frac{u}{n}} \right) + \frac{{\overline A }}{A} \cdot {H^2} \cdot c \cdot \Delta \left( {\frac{1}{n}} \right)} \right\}} ,
\end{array}} \right.
\label{eq:7}
\end{equation}
where $A$ represents the Snell invariant of the marginal ray,  $\overline{A}$\ denotes the Snell invariant of the chief ray, and $H$ is the Lagrange invariant of the system. They can be calculated as:
\begin{equation}
\left\{ \begin{array}{l}
A = n\left( {yc + u} \right)\\
\bar A = n\left( {\bar yc + \bar u} \right)\\
\Delta \left( {\frac{u}{n}} \right){\rm{ = }}\frac{u'}{n'} - \frac{{u}}{{n}}\\
\Delta \left( {\frac{1}{n}} \right){\rm{ = }}\frac{1}{n'} - \frac{1}{{n}}\\
H = n\bar uy - nu\bar y .
\end{array} \right.
\label{eq:8}
\end{equation}

In this paper, our main concerns are about the spherical aberration, coma, astigmatism and Petzval curvature of TMA. The distortion can be corrected by the specific image processing algorithm, so it is not considered in the design process. From Eqs. (\ref{eq:3}) - (\ref{eq:8}), the Seidel coefficients can be expressed as the functions of $y_1$, $\overline{y_1}$\ and $r_i$ ( $i$ =1, 2, 3), $t_i$ ( $i$ =1, 2, 3, 4) as Eq. (\ref{eq:9}). It can be seen that rays' heights $y_1$ and $\overline{y_1}$\ in each aberration coefficient only control the scales of the aberrations and do not affect the aberrations distribution.

\begin{figure*}[ht]
\newcounter{TempEqCnt}     
\setcounter{TempEqCnt}{\value{equation}} 
\setcounter{equation}{8}  
\begin{equation} 
\left\{ {\begin{array}{*{20}{l}}
\begin{array}{l}
{S_{I}} ={\frac{2y_1^4}{{r_1^3}}} [ {-1+ \frac{1}{{r_1r_2^3}}{( {{r_1} - 2{t_2}} )^2}{( {2{r_2} - {r_1} + 2{t_2}} )^2} - \frac{1}{{r_1r_2^4r_3^3}}
 ( {{r_1}{r_2} - 2{r_1}{t_3} - 2{r_2}{t_2} + 2{r_2}{t_3} + 4{t_2}{t_3}} )^2
( {{r_1}{r_2}  } }\\
\ \qquad{ {- 2{r_1}{r_3} + 2{r_2}{r_3} - 2{r_1}{t_3} -2{r_2}{t_2}+ 2{r_2}{t_3} + 4{r_3}{t_2} + 4{t_2}{t_3}})^2} ]
\end{array}\\
\begin{array}{l}
{S_{II}} = {\frac{2y_1^3{{\bar y}_1}}{{r_1^3{t_1}}}}  [ - ( {{r_1} + {t_1}} )+ \frac{1}{{r_1r_2^3}}{ ( {{r_1} - 2{t_2}}  )^2} ( {2{r_2} - {r_1} + 2{t_2}}  )
 ( {{r_1}{r_2} - {r_1}{t_1} + {r_1}{t_2} + 2{r_2}{t_1} + 2{t_1}{t_2}}  )
 - \frac{1}{{r_1r_2^4r_3^3}}\\
 \ \qquad \cdot {( {r_1}{r_2} - 2{r_1}{t_3} - 2{r_2}{t_2} + 2{r_2}{t_3} + 4{t_2}{t_3}} )^2
 ( {{r_1}{r_2} - 2{r_1}{r_3} + 2{r_2}{r_3} - 2{r_1}{t_3} - 2{r_2}{t_2}+ 2{r_2}{t_3}+ 4{r_3}{t_2} }\\
\ \qquad  {+ 4{t_2}{t_3}} )( {{r_1}{r_2}{r_3} + {r_1}{r_2}{t_1} - {r_1}{r_2}{t_2} - 2{r_1}{r_3}{t_1} + {r_1}{r_2}{t_3}+ 2{r_1}{r_3}{t_2}+ 2{r_2}{r_3}{t_1}- 2{r_1}{t_1}{t_3} - 2{r_2}{t_1}{t_2}}\\
\ \qquad  { + 2{r_1}{t_2}{t_3}+ 2{r_2}{t_1}{t_3} + 4{r_3}{t_1}{t_2} + 4{t_1}{t_2}{t_3} } )]
\end{array}\\
\begin{array}{l}
{S_{III}} = {\frac{2y_1^2\bar y_1^2}{{r_1^3t_1^2}} }  [ -{ ( {{r_1} + {t_1}}  )^2} + \frac{1}{{r_1r_2^3}}{ ( {{r_1} - 2{t_2}}  )^2}{ ( {{r_1}{r_2} - {r_1}{t_1} + {r_1}{t_2} + 2{r_2}{t_1} + 2{t_1}{t_2}}  )^2} 
 - \frac{1}{{r_1r_2^4r_3^3}} ( {{r_1}{r_2} - 2{r_1}{t_3}}\\
 \ \qquad {- 2{r_2}{t_2} + 2{r_2}{t_3} + 4{t_2}{t_3}}  )^2  ( {{r_1}{r_2}{r_3} + {r_1}{r_2}{t_1} - {r_1}{r_2}{t_2} - 2{r_1}{r_3}{t_1} + {r_1}{r_2}{t_3} + 2{r_1}{r_3}{t_2} + 2{r_2}{r_3}{t_1} - 2{r_1}{t_1}{t_3} }\\
\ \qquad {- 2{r_2}{t_1}{t_2} + 2{r_1}{t_2}{t_3} + 2{r_2}{t_1}{t_3}+ 4{r_3}{t_1}{t_2} + 4{t_1}{t_2}{t_3}} )^2 ]
\end{array}\\
\begin{array}{l}
{S_{IV}} =   \frac{2{y_1^2}\bar y_1^2}{{{r_1}{r_2}{r_3}t_1^2}} ( {{r_1}{r_2} - {r_1}{r_3} + {r_2}{r_3}}  ).\\
\end{array}
\end{array}} \right. 
\label{eq:9}
\end{equation}
\end{figure*}
\setcounter{equation}{\value{TempEqCnt}}

 For a general design process, a specific Seidel coefficient is expected to be zero for different design requirements. For example, if a flat image field is desired, we make $S_{IV}$ =0, and then solve the equations to acquire the corresponding relationship among the parameters. By setting reasonable initial values of $y_1$, $\overline{y_1}$\ , $t_1$ and $t_3$, the structural parameters that minimize these four aberrations and satisfy $y_4$ = 0 can be determined, thus obtaining the initial optical structure.

\section{Design and optimization process}
\label{sec:3}
\subsection{Initial Coaxial System Design}
For this design, we set the initial values of $y_1$ = 80mm, $\overline{y}_1$\ = 127mm, $t_1$ = -180mm and $t_3$ = 4000mm to make the Seidel coefficients target to zero. The initial structural parameters of the coaxial system, including the curvature radii of three mirrors and the distances between them, are derived based on the coaxial system design method described above. These parameters are listed in Table \ref{tab:2}. All mirrors are initially designed with spherical surfaces with conic coefficients. The virtual stop, namely the entrance pupil, is located behind of the PM. Since the real exit pupil is the conjugate image of the entrance pupil, we will remove the stop at the entrance pupil and set a scanning mirror at the exit pupil position as the new stop later on. In the following design process, the commercial software ZEMAX \cite{zemax} will be employed for the further design and optimization process. 

\begin{table}[htbp]
\centering
\caption{ Coaxial Initial Structure Parameters}
\begin{tabular}{ccccc}
\hline
Surface & Radius(mm) &  Distance(mm) & Conic &  Size(mm) \\
\hline
Stop & - & -180 & - & $\Phi$160\\
PM & 1320.104 &-3065.241 & 0.904 & $\Phi$254\\
SM & 2741.618 & 4000.000 & -0.006 &  $\Phi$1920\\
TM & -2512.183 & -4000.000 & -0.118 &  $\Phi$1262\\
Image & - & - & - & $\Phi$438\\
\hline
\end{tabular}
\label{tab:2}
\end{table}

\subsection{ Off-axis TMA Design}

\begin{figure}[ht!]
\centering
\includegraphics[width=.8\textwidth]{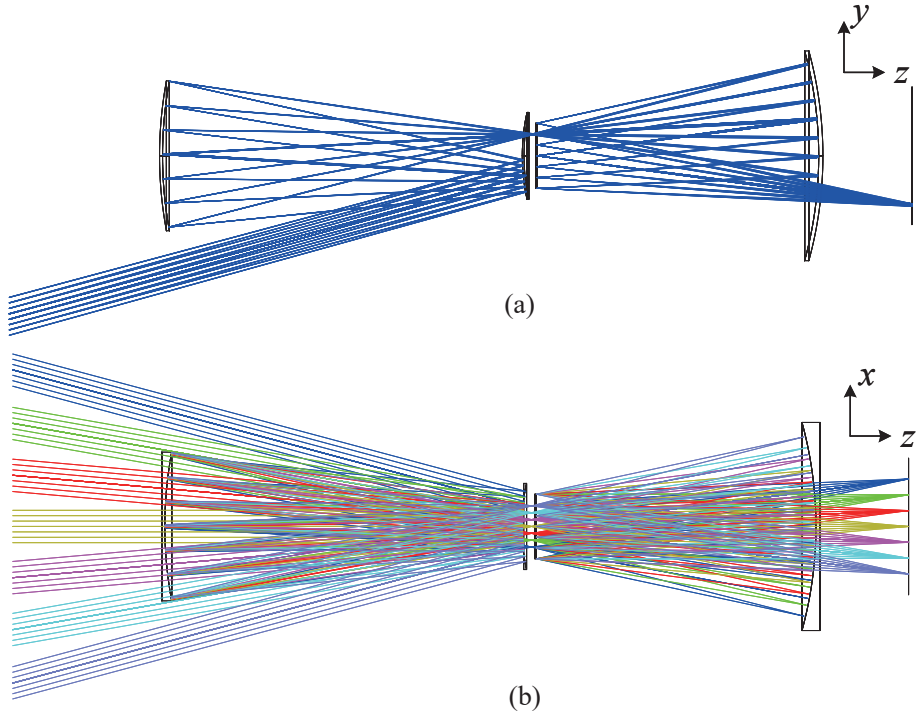}
\caption{The initial off-axis structure: (a) side view; (b) top view.
}
\label{fig:4}
\end{figure}

After obtaining the initial coaxial structure, we set the FOV as an off-axis configuration in the $y$ direction to eliminate the mirror occlusion. Specifically, the field angle of 0.5° in the $y$ direction is rotated 15° counterclockwise and  the optical system is symmetric about the $y$ - $z$ plane. In this way, the FOV in $y$ direction is the range from 15$^{\circ}$ to 15.5$^{\circ}$, and the FOV in $x$ direction is the range from -15$^{\circ}$ to 15$^{\circ}$. The obtained optical system  with the off-axis FOV is shown in Fig. \ref{fig:4}. We introduce a folding  plane mirror between TM and the image plane for better visualization of the optical path. In the actual optimization process, this folding mirror is not considered.

\begin{figure}[ht!]
\centering
\includegraphics[width=.8\textwidth]{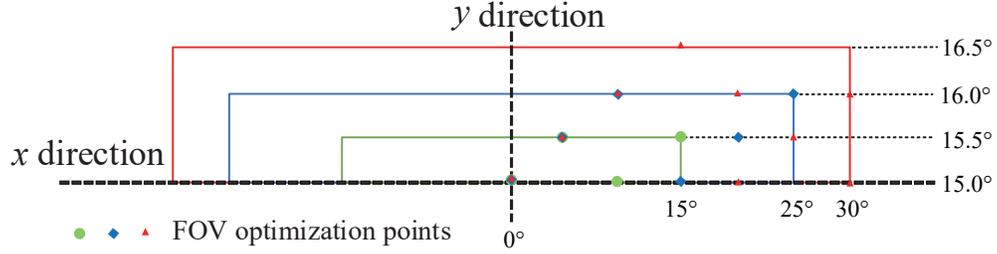}
\caption{Schematic diagram of gradually increasing FOV.}
\label{fig:5}
\end{figure}

In the subsequent stage of expanding the FOV of system, the image quality is the most critical concern to pay attention to, followed by the problem of mirror occlusion. When the image quality is optimized to the best state, we expand FOV in $x$ to $\pm50^{\circ}$, while increasing the FOV in $y$ direction as a new range from 15$^{\circ}$ to 16$^{\circ}$. At the same time, we set a higher weight coefficient of the margin FOV compared with that of the central FOV to compensate for the poor edge imaging quality. Repeating the above optimization procedure, we finally obtained a phased design with a FOV of 60$^{\circ}$×1.5$^{\circ}$. The whole expanding process of FOV is shown in Fig. \ref{fig:5}.

\subsection{Scanning Mirror Design}
For an optical system with a large FOV, the image plane is always so large that is not conducive to optical observation and system construction. This problem is usually solved by adding a scanning mirror along the path of light propagation. However, for the system we obtained above, the exit pupil distance varies greatly with different sub-FOVs. In other words, different chief rays correspond to different optical paths due to the rotation of the scanning mirror and makes defocusing occur. Alternatively, the multi-configuration structure also has the potential to solve this problem. However, if the multi-configuration structure approach is adopted before the final optimization, it will result in huge complexity in the following optimization process due to the numerous variables. 

\begin{figure}[ht!]
\centering
\includegraphics[width=.8\textwidth]{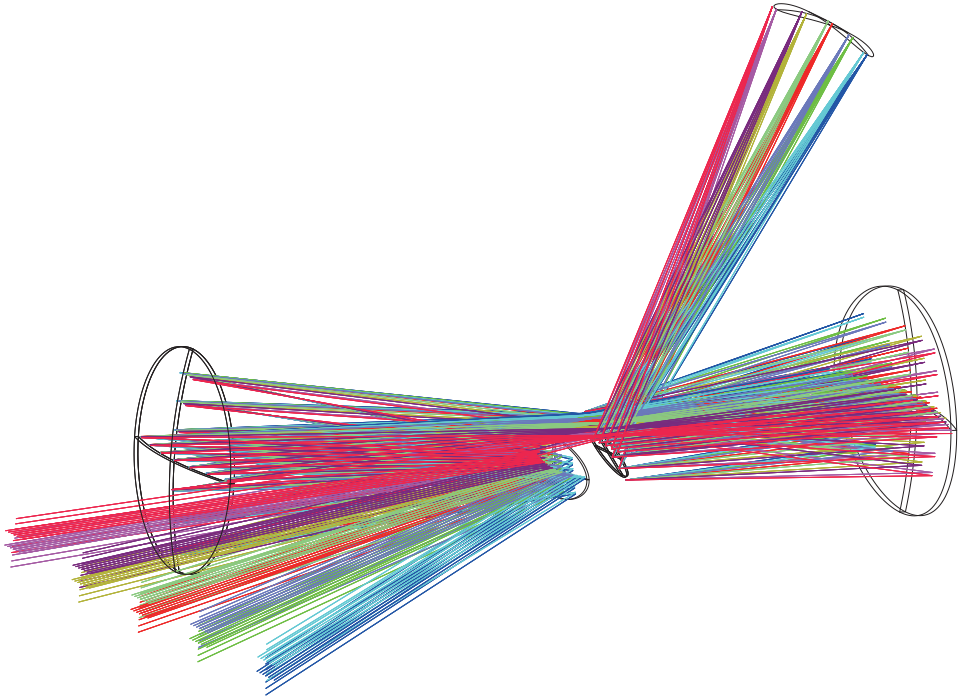}
\caption{Schematic diagram of the image surface as a curved surface.}
\label{fig:6}
\end{figure}

To solve this problem, we propose a novel method to control the exit pupil distance. Since the size of the scanning mirror is much smaller compared with the distance between the image plane and the scanning mirror, the image plane can be set as a curved surface under approximate conditions. We set the radius of curvature of the curved surface to be equal to the exit pupil distance of the chief ray in the central FOV. So, the exit pupil distance of each FOV can be roughly regarded as equal. Based on this assumption, we set the scanning mirror at the 
position of the exit pupil, and the state is fixed temporarily. Now, there is only one configuration with a large FOV. Then, we optimize the surfaces of PM, SM, and TM and the distances between them. After the optimization, we obtain the optical system that meets the preset requirements as shown in Fig. \ref{fig:6}. 

However, it is not appropriate to set the image surface as a curved one in practical application and  the caliber of TM reaches 1250 mm and the diameter of the image plane is 860 mm, they are so large that bring a great challenge to the manufacturing as well as the alignment. So, in the following optimization, we will remove the radius of curvature of the image surface and combine the deflection angle of the scanning mirror to redistribute the full FOV to multi-configuration design in the $x$ direction. The scanning mirror with $\pm9^{\circ}$ rotation is added to the multi-configuration structure as listed in Table \ref{tab:3}. In the subsequent optimization process, the group of mirrors before the scanning mirror is fixed and does not participate in the optimization. As the entire optical system is symmetrical about the $y$ - $z$ plane in terms of mirrors and structural settings, only six half FOVs in the $x$ direction will be analyzed, which are corresponding to six configurations in ZEMAX. For the $y$ direction, the angle of $15^{\circ}$, $15.5^{\circ}$, $16^{\circ}$ and $16.5^{\circ}$ are considered in each configuration. 

\begin{table*}[htbp]
\centering
\caption{ Multi-configuration Settings}
\begin{tabular}{ccccccc}
\hline
Variable & Config. 1 &  Config. 2 & Config. 3 &  Config. 4 & Config. 5 & Config. 6 \\
\hline
XFIE 1 & 0° & 10°  & 15° & 20° & 25° & 30°\\
XFIE 2 & 0° & 10°  & 15° & 20° & 25° & 30°\\
XFIE 3 & 0° & 10°  & 15° & 20° & 25° & 30°\\
XFIE 4 & 0° & 10°  & 15° & 20° & 25° & 30°\\
Scanning mirror rotation angle & 0° & 3° & 4.5° & 6° & 7.5° & 9°\\
\hline
\end{tabular}
\label{tab:3}
\end{table*}

The approximation in the previous process is indeed beneficial to the optimization process, but the optimized structure is not completely equivalent to the structure obtained by adding multi-configuration at the end, which will inevitably introduce residual aberrations. For an industrial application, a deformable mirror can be employed to be installed behind the scanning mirror to improve the image quality.

\section{Design results and sensitivity analysis}
\label{sec:4}
\subsection{Design results}

\begin{figure}[ht!]
\centering
\includegraphics[width=.8\textwidth]{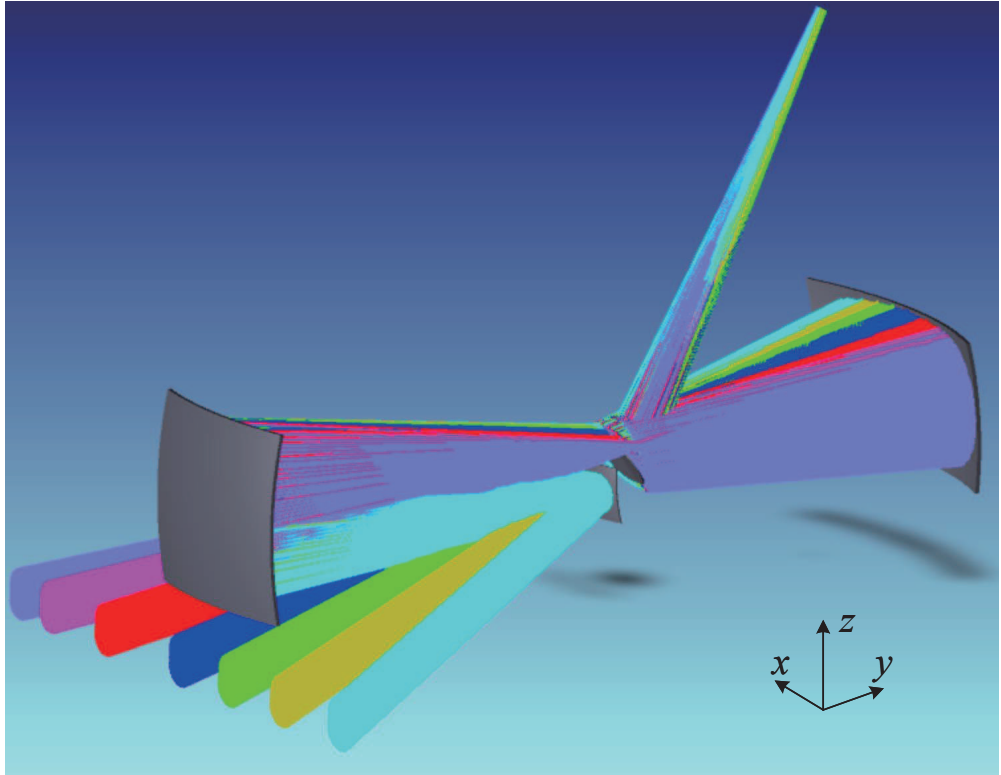}
\caption{The optical layout of the final obtained telescope.}
\label{fig:7}
\end{figure}

The final optimized optical system is shown in Fig. \ref{fig:7}. To make a relatively compact structure, the apertures of the PM, SM and TM are restricted to rectangles. EFL of this telescope equals 876mm, and other optical properties meet the initial design specifications as listed in Table \ref{tab:1}. The structural parameters are detailed in Table \ref{tab:4} and Table \ref{tab:5}. PM, TM and SM are all even-order aspheric surfaces with eighth-order coefficients, and only fourth-order, sixth-order and eighth-order coefficients are used.

\begin{table*}[!ht]
\centering
\caption{ Parameters of Final Structure}
\begin{tabular}{cccccc}
\hline
Surface & Radius&  Distance & Conic &  Size   & tilted about $x$ \\
\hline
PM & 803.622mm &-165.2630mm & 0.687 & 250mm*110mm & $0^{\circ}$\\
SM & 1954.405mm & 3207.541mm & -0.075 & 420mm*380mm & $0^{\circ}$\\
TM & -1894.321mm & -1328.788mm & -0.094 & 600mm*300mm & $0^{\circ}$\\
Coordinate Break & Infinity & 0mm & - & - & $-35^{\circ}$\\
Scanning mirror & Infinity & 0mm & 0 & $\Phi$110mm & $0^{\circ}$\\
Coordinate Break & Infinity & 0mm & - & - & $35^{\circ}$\\
Coordinate Break & Infinity & 1893.190mm & - & - & $-63^{\circ}$\\
Image & - & - & - & $\Phi$22mm & $0^{\circ}$\\
\hline
\end{tabular}
\label{tab:4}
\end{table*}

\begin{table}[ht]
\centering
\caption{ Detailed Even-order Aspheric Coefficients of Mirrors}
\begin{tabular}{cccc}
\hline
Surface & 4th order &  4th order &4th order\\
\hline
PM & 1.218e-10 & -1.817e-16 & 5.420e-21 \\
SM & -1.552e-12 & -9.185e-19 & 1.750e-24\\
TM & -2.817e-12 & 4.801e-19 & -2.076e-24 \\
\hline
\end{tabular}
\label{tab:5}
\end{table}

\begin{figure}[ht!]
\centering
\includegraphics[width=.8\textwidth]{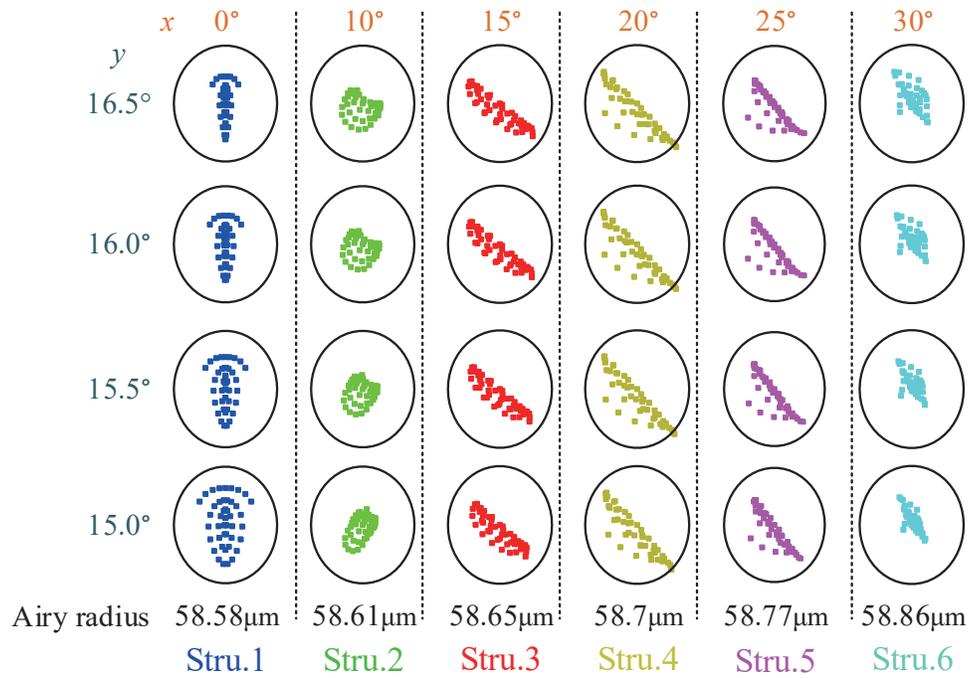}
\caption{The spot diagram of the final obtained telescope.}
\label{fig:8}
\end{figure}

The standard spot diagrams on the image plane of all configurations are exhibited in Fig. \ref{fig:8}. To facilitate the comparison of the relationship with the Airy disk, the size and position of the Airy disks are also provided. It is obvious that all RMS radii are less than 35$\upmu$m which is much smaller than the Airy disk radius of 59$\upmu$m.

\subsection{Tolerance Analysis}
As mentioned in Section 1, the off-axis systems are more difficult to manufacture and align than the general coaxial systems due to the asymmetric aberrations caused by the non-rotationally symmetric elements. To demonstrate the instrumentation feasibility of the obtained telescope, a brief tolerance analysis is provided. The tolerance distribution of each component is listed in Table \ref{tab:6}.
\begin{table}[ht]
\centering
\caption{ Tolerance Setting in ZEMAX}
\begin{tabular}{ccc}
\hline
Surface & Tolerances & Value\\
\hline
All & Radius & ±10$\upmu$m\\
& Thickness & ±50$\upmu$m\\
PM & Decenter in $x$ & ±50$\upmu$m\\
& Decenter in $y$ & ±50$\upmu$m\\
& Tilt in $x$ & 0.01°\\
& Tilt in $y$ & 0.01°\\
SM & Decenter in $x$ & ±20$\upmu$m\\
& Decenter in $y$ & ±20$\upmu$m\\
& Tilt in $x$ & 0.01°\\
& Tilt in $y$ & 0.01°\\
TM & Decenter in $x$ & ±50$\upmu$m\\
& Decenter in $y$ & ±50$\upmu$m\\
& Tilt in $x$ & 0.02°\\
& Tilt in $y$ & 0.02°\\
Scanning mirror & Decenter in $x$ & ±50$\upmu$m\\
& Decenter in $y$ & ±50$\upmu$m\\
& Tilt in $x$ & 0.04°\\
& Tilt in $y$ & 0.04°\\
\hline
\end{tabular}
\label{tab:6}
 \end{table}

\begin{figure}[ht!]
\centering
\includegraphics[width=.8\textwidth]{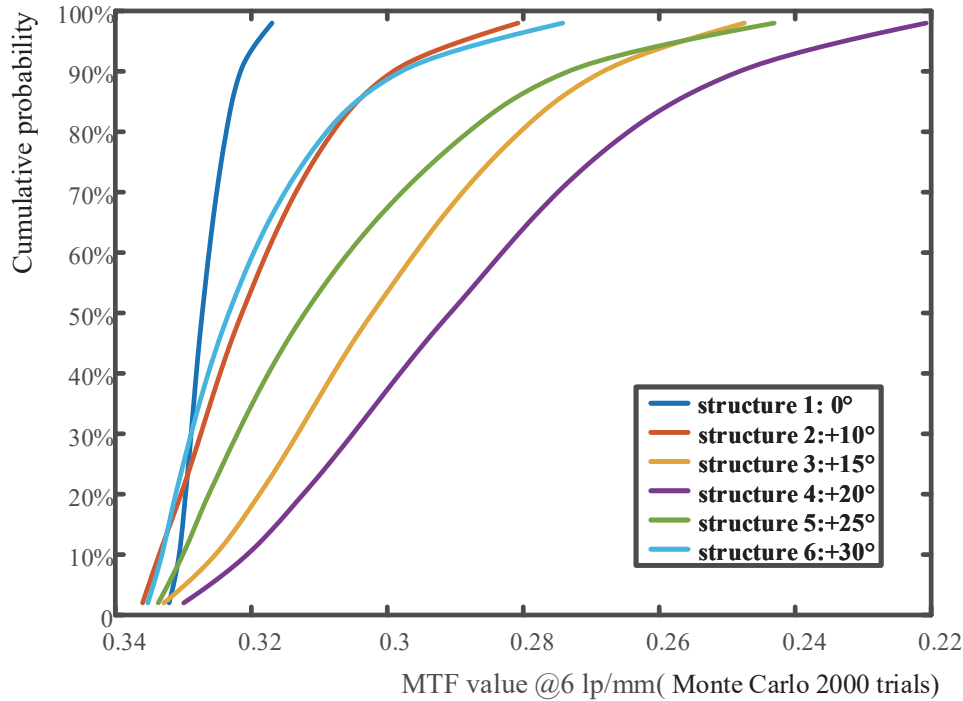}
\caption{Cumulative probability change of MTF at 6 lp/mm}
\label{fig:9}
\end{figure}

We have performed 2000 Monte Carlo sensitivity analysis to predict the optical performance. The average diffraction MTF at 6 lp/mm (60\% of the Nyquist frequency of 10 lp/mm) is selected as the criteria for the evaluation of optical performance. The overall analysis results are shown in Fig. \ref{fig:9} and listed in Table \ref{tab:7}. The image quality from $15^{\circ}$ to $25^{\circ}$ FOV is poorer than that of the center FOV (0$^{\circ}$) and the edge FOV (30$^{\circ}$), which coincides with the result of the standard spot diagram. Nevertheless, 90\% MTF is larger than 0.24, 80\% MTF is larger than 0.26 and 50\% MTF is larger than 0.29 for all FOV in different structures. As a consequence, the obtained optical system has been proved to have good instrumentation feasibility.

\begin{table}[!ht]
\centering
\caption{ 2000 Trails Monte Carlo Tolerance Analysis Probability Results of MTF at 6lp/mm}

\begin{tabular}{ccccccc}
\hline
Configurations & 98\% & 90\% & 80\%  & 50\% \\
\hline
Config. 1 & 0.3170 & 0.3216 & 0.3237 & 0.3272\\
Config. 2 & 0.2808 & 0.2991 & 0.3080 & 0.3214\\
Config. 3 & 0.2475 & 0.2683 & 0.2805 & 0.3022\\
Config. 4 & 0.2207 & 0.2482 & 0.2645 & 0.2907\\
Config. 5 & 0.2431 & 0.2734 & 0.2881 & 0.3262\\
Config. 6 & 0.2742 & 0.2980 & 0.3089 & 0.3233\\
\hline
\end{tabular}
\label{tab:7}
\end{table}

\section{Conclusion}
\label{sec:5}

In this paper, a high throughput telescope based on scanning off-axis TMA system has been successfully designed. We provide an innovative
solution to obtain a large rectangle full FOV by employing a scanning mirror at the real exit pupil. Combined with the progressive optimization method, we finally obtain a telescope with a large FOV of 60$^{\circ}$×1.5$^{\circ}$, an F number of 6, a focal length of 876mm, and the  optical resolution reaches the diffraction limit. We have also provided a brief tolerance analysis to demonstrate the robustness of the proposed design method. The mirrors are the aspheric surfaces with eighth-order coefficients in this design. If more complex surfaces are adopted, such as the extended polynomial surfaces, a larger FOV and better image quality would be achieved theoretically. And this will be researched in the follow-up work.

\bigskip

\noindent\textbf{Funding.} National Natural Science Foundation of China (61805088); Science, Technology, and Innovation Commission of Shenzhen Municipality (JCYJ20190809100811375); Key Research and Development Program of Hubei Province (2020BAB121); Fundamental Research Funds for the Central Universities (2019kfyXKJC040); Innovation Fund of WNLO.

\medskip

\noindent\textbf{Disclosures.} The authors declare no conflicts of interest.

\end{document}